\documentclass[letter]{aa}

\usepackage{graphicx}
\usepackage{txfonts}
\usepackage{hyperref}
\usepackage{graphicx}
\usepackage{bm}
\usepackage{amsmath}
\usepackage{amssymb}
\usepackage{amsfonts}
\usepackage{hyperref}
\usepackage{adjustbox}
\usepackage{natbib}
\usepackage{soul}

\bibpunct{(}{)}{;}{a}{}{,} 
\usepackage{bbold}
\hypersetup{colorlinks=true,citecolor=blue,linkcolor=blue,urlcolor=blue}
\usepackage{tabularx}
\usepackage{xspace}

\newcommand*{\Euclid}{\textit{Euclid}\xspace}

\definecolor{orange}{rgb}{1,0.5,0}
\definecolor{darkorange}{rgb}{0.69,0.33,0.13}
\definecolor{fidcol}{rgb}{0.7,0,0}     
\definecolor{mkcol}{rgb}{0.5,0,0.5}
\definecolor{mmcol}{rgb}{0.7,0.17,0.31}
\definecolor{dscol}{rgb}{0.6,0.1,0.2}
\definecolor{mccol}{rgb}{0.2,0.4,0.6}
\definecolor{darkgreen}{rgb}{0.05,0.5,0.06}
\definecolor{carnelian}{rgb}{0.7, 0.11, 0.11}
\definecolor{whatever}{rgb}{0.7,0.5,0.2}

\begin{document} 

\title{Starlet $\ell_1$-norm for weak lensing cosmology}

\author{Virginia Ajani
          \inst{1}
          \and Jean-Luc Starck
          \inst{1}
          \and Valeria Pettorino
          \inst{1}
          }
    
\institute{AIM, CEA, CNRS, Universit{\'e} Paris-Saclay, Universit{\'e} de Paris, 
             Sorbonne Paris Cit{\'e}, F-91191 Gif-sur-Yvette, France\\
              \email{virginia.ajani@cea.fr}}
 
 \date{Received 25/11/2020; Accepted 22/12/2020}            

\abstract{We present a new summary statistic for weak lensing observables, higher than second order, suitable for extracting non-Gaussian cosmological information and inferring cosmological parameters. We name this statistic the `starlet $\ell_1$-norm' as it is computed via the sum of the absolute values of the starlet (wavelet) decomposition coefficients of a weak lensing map. In comparison to the state-of-the-art higher-order statistics -- weak lensing peak counts and minimum counts, or the combination of the two -- the $\ell_1$-norm provides a fast multi-scale calculation of the full void and peak distribution, avoiding the problem of defining what a peak is and what a void is: The  $\ell_1$-norm carries the information encoded in all pixels of the map, not just the ones in local maxima and minima. We show its potential by applying it to the weak lensing convergence maps provided by the \url{MassiveNus} simulations to get constraints on the sum of neutrino masses, the matter density parameter, and the amplitude of the primordial power spectrum. We find that, in an ideal setting without further systematics, the starlet $\ell_1$-norm remarkably outperforms commonly used summary statistics, such as the power spectrum or the combination of peak and void counts, in terms of constraining power, representing a promising new unified framework to simultaneously account for the information encoded in peak counts and voids. We find that the starlet $\ell_1$-norm outperforms the power spectrum by $72\%$ on M$_{\nu}$, $60\%$ on $\Omega_{\rm m}$, and $75\%$ on $A_s$ for the \Euclid-like setting considered; it also improves upon the state-of-the-art combination of peaks and voids for a single smoothing scale by $24\%$ on M$_{\nu}$, $50\%$ on $\Omega_{\rm m}$, and $24\%$ on $A_s$.}

\maketitle
%

\section{Introduction}

Weak gravitational lensing by the large-scale structure represents a powerful tool for estimating cosmological parameters. Past, present, and future cosmological surveys, such as the Canada-France-Hawaii Telescope Lensing Survey (CFHTLenS) \citep{Heymans_2012}, the Kilo-Degree Survey (KiDS) \citep{heymans2020kids1000}, the Dark Energy Survey (DES) \citep{2019PhRvD..99l3505A, to2020dark}, Hyper SuprimeCam (HSC) \citep{2017AAS...22922602M}, \Euclid \citep{2011arXiv1110.3193L}, and the Rubin Observatory \citep{2009arXiv0912.0201L},
consider it as one of the main physical probes for investigating unsolved questions in current cosmology, such as: what the properties of the dark components of the universe are, what the origin of its accelerated expansion is \citep{1998AJ....116.1009R,1999ApJ...517..565P}, and what the sum of neutrino masses is \citep{2012arXiv1212.6154L}. It is very well known that, in the context of weak lensing, second-order statistics as the two-point correlation function or its Fourier transform (the power spectrum) do not capture the non-Gaussian information encoded in the non-linear features of weak lensing data \citep{2013PhR...530...87W}. This has motivated the introduction of several higher-order statistics, such as Minkowski functionals \citep{2012PhRvD..85j3513K,2015PhRvD..91j3511P,  2019PhRvD..99d3534V, 2019JCAP...06..019M, 2020A&A...633A..71P}, higher-order moments \citep{2016PhRvD..94f3534P, 2018PhRvD..97b3519V, 2018A&A...619A..38P, 2018MNRAS.475.3165C, 2020MNRAS.498.4060G}, the bispectrum \citep{2004MNRAS.348..897T, 2017PhRvD..96b3528C,2019JCAP...05..043C},
peak counts \citep{1999MNRAS.302..821K,2010PhRvD..81d3519K, 2010MNRAS.402.1049D,2011MNRAS.416.2527M,2012MNRAS.423..983P,2012MNRAS.425.2287H,2012MNRAS.426.2870H,2012MNRAS.423.1711M,2013MNRAS.432.1338M, 2015A&A...581A.101M,2015A&A...583A..70L,2018MNRAS.478.5436G,2018MNRAS.474..712M,2018A&A...619A..38P, 2019PhRvD..99f3527L}, the scattering  transform \citep{2020MNRAS.tmp.2970C}, wavelet phase harmonic statistics \citep{PhysRevD.102.103506}, and machine learning-based methods \citep{2018PhRvD..98l3518F, 2019PhRvD.100b3508P,2018PhRvD..97j3515G,10.5555/1162254, 2019MNRAS.490.1843R, 2019arXiv191112890S}, to account for non-Gaussian information in cosmological analysis. Focusing on peak counts, it has been shown that this statistic is particularly powerful in breaking degeneracy between the standard model and fifth forces in the dark sector \citep{2018A&A...619A..38P} as well as in constraining cosmological parameters when employed in a multi-scale setting \citep{2015PhRvD..91f3507L, 2016A&A...593A..88L, 2018JCAP...10..051F, PhysRevD.102.103531, 2020arXiv200612506Z}. In particular, \cite{PhysRevD.102.103531} have shown that multi-scale peak counts significantly outperform the weak lensing power spectrum, improving the constraints on the sum of neutrino masses $\sum m_{\nu} \equiv M_{\nu}$ by 63\% when using a starlet filter; multi-scale peak counts were also shown to be so constraining that the addition of the power spectrum does not further improve constraints. A very interesting feature of multi-scale peaks, when they are obtained using the starlet transform \citep{Starck2007}, is the behaviour of the covariance matrix that tends to encode all information in its diagonal elements.

Another weak lensing probe of large-scale structure is represented by
cosmic voids, namely under-dense regions of the large-scale matter field \citep{2008MNRAS.387..933C, 2019BAAS...51c..40P}. Local minima of weak lensing convergence maps, namely pixels with values smaller than their eight neighbouring  pixels, have been proposed as tracers of the matter distribution voids to infer cosmological parameters, both in a mono-scale setting \citep{2020MNRAS.495.2531C, 2020arXiv201007376M, 2020arXiv201011954D} and in a multi-scale setting \citep{2020arXiv200612506Z}.
More specifically, \cite{2020MNRAS.495.2531C} found that lensing minima alone are slightly less constraining than the peaks alone and, in agreement with \cite{2020arXiv201007376M} and \cite{2020arXiv200612506Z}, that the combination of the two statistics produces significantly tighter constraints than the power spectrum.

In this paper, we propose, for the first time, using the $\ell_1$-norm of wavelet coefficients of weak lensing convergence maps.\ We show that it provides a unified framework for a joint multi-scale peak and void analysis and that it takes into account the information encoded in all pixels of the map. 

This letter is structured as follows: In Sect. \ref{sec: L1_norm}, we define the $\ell_1$-norm. In Sect. \ref{sec: Analysis}, we specify the details for the analysis, defining the summary statistics employed and the corresponding choice of filters and binning. We show how we compute the covariance matrix, show how we build the likelihood, show the posterior distributions, and define the estimators used to quantify our results. In Sect. \ref{sec: Results}, we describe our results, and we draw our conclusions in Sect. \ref{sec: Conclusions}. We dedicate  \autoref{AppendixA} to background definitions in weak lensing and to the details of the simulations. In \autoref{AppendixB}, we discuss features of covariance matrices.

\section{Towards the starlet $\ell_1$-norm }\label{sec: L1_norm}
\subsection{Starlet peaks}
Multi-scale peak counts can be derived using either a set of Gaussian kernels of different sizes or a wavelet decomposition, such as the starlet transform \citep{PhysRevD.102.103531}. 
The starlet transform decomposes a convergence map $\kappa$ of size $N \times N$
into a set ${\cal W} = \lbrace w_1, ... , w_{j_{max}}, c_J \rbrace$ of $J = j_{max} + 1$ bands of the same size, where $j_{max}$ is the number of wavelet scales considered,  $c_J$ is
the coarse scale, namely a very smoothed version of the original image $\kappa$, 
and $w_j$ are the wavelet bands at scale $2^j$ pixels. A complete description and derivation of the starlet transform algorithm can be found in  \cite{Starck2007,starck:book15}. 
It was also shown in \citet{LPS.2012} that 
the starlet transform can be interpreted as a fast multi-scale aperture mass decomposition where
the aperture mass kernels have a compact support and are compensated functions (namely, the kernel integral is null). Starlet peaks are then derived by considering $n$ bins with 
bin edges given by the minimum and maximum values of each band in ${\cal W}$.
An interesting advantage of such an approach is that each wavelet band covers a different frequency range, which leads to an almost diagonal peak count covariance matrix 
\citep{PhysRevD.102.103531}. This is not the case when a standard multi-scale Gaussian analysis is applied on the convergence map.

\subsection{Starlet extrema}
As mentioned in the introduction, cosmic void analysis is an alternative to peak analysis for studying convergence maps, and the combination of the two improves the constraints on cosmological parameters.
It is interesting to notice how a starlet decomposition can naturally include a multi-scale void analysis. Instead of extracting maxima (peaks) in each band, we can also extract minima
(pixels  with  values smaller than  their   eight   neighbours), and a joint peak-void multi-scale is therefore obtained by extracting wavelet coefficient extrema (minima+maxima). 
The starlet decomposition therefore provides a very natural framework for a joint multi-scale peak and void analysis. 
\subsection{Starlet $\ell_1$-norm}
A particularity of peak and void statistics is that only a few pixels are considered,  
while other high-order statistics, such as bispectrum or Minkowski functionals, use all the pixels.
In a starlet framework, we should emphasise that starlet peaks have mainly positive values and starlet voids negative values due to the property of the wavelet function. So instead of counting the number of peaks or voids in a given bin $i$ 
defined by two values, $B_i$ and $B_{i+1}$, we could 
take the sum of all wavelet coefficients with an amplitude between $B_i$ and $B_{i+1}$.
If $B_i$ and $B_{i+1}$ are positive, this corresponds to the definition of the set of coefficients ${\cal S}_{j,i}$ at scale $j$ and in bin $i$ such that ${\cal S}_{j,i} = \lbrace w_{j,k} / B_i <  w_{j,k}  < B_{i+1} \rbrace $, where $k$ is the pixel index.\ We can then compute the sum 
$\sum_{u=1}^{\#coef({\cal S}_{j,i})}  {\cal S}_{j,i} [u] $. This can be generalised to positive and negative bins using:
\begin{equation}
    l_1^{j,i} = \displaystyle \sum_{u=1}^{\#coef({\cal S}_{j,i})}  \mid {\cal S}_{j,i} [u]  \mid =  || {\cal S}_{j,i} ||_1
    \label{eq: L1_norm_starlet}
,\end{equation}
where $ || . ||_1$ is the standard $\ell_1$-norm (i.e. $ || x ||_1 = \sum_k | x_k | $) and
the index $u$ runs from 1 to the number of pixels in a given bin $i$ at scale $j$ (i.e. $\#coef({\cal S}_{j,i})$). 
The quantity $l_1^{j,i} $ defined in Equation \eqref{eq: L1_norm_starlet}
is nothing more than the $\ell_1$-norm of the binned pixel values of the starlet coefficients of the original image $\kappa$ map. In the following, we will name $S_{\ell_1}$, the starlet $\ell_1$-norm, as the set $S_{\ell_1}$ of all $l_1^{j,i}$ numbers obtained from the different scales $j$ and bins $i$. 

This approach enables us to extract the information encoded in the absolute value of all pixels in the map instead of characterising it only by selecting local minima or maxima. 
An interesting advantage is that it avoids the open issue of how to define a void \citep{2008MNRAS.387..933C}. It is interesting to notice that this $S_{\ell_1}$ statistic is also directly related to the density probability function of the starlet coefficients at different scales. 

\section{Experiment}\label{sec: Analysis}
\subsection{Summary statistics}\label{subsec: summary_statistics}
We provided constraints on the sum of neutrino masses $M_{\nu}$,  on the matter density parameter $\Omega_{\rm m}$, and  on  the power  spectrum  amplitude $A_{\rm s}$ by employing five different statistics applied to the \url{MassiveNus}\footnote{Further details on the simulations are illustrated in \autoref{AppendixA}.} simulations \citep{2018JCAP...03..049L}. Three of them are 
  statistics that are used in weak lensing studies: the power spectrum, 
  combined mono-scale peak and void counts, and multi-scale peak count (using the starlet transform). The two others are the starlet extrema and the starlet $\ell_1$-norm $S_{\ell_1}$ that we introduced in the previous section. 
  
We computed the summary statistics on maps of the signal-to-noise field, where we define the signal to noise as the ratio between the noisy convergence $\kappa$ convolved with the filter $\mathcal{W}(\theta_\mathrm{ker})$ over the smoothed standard deviation of the noise for each realisation per redshift: 
\begin{equation}\label{eq:singal_to_noise}
S/N \equiv \frac{(\mathcal{W} \ast \kappa )(\theta_\mathrm{ker})}{\sigma_{n}^\mathrm{filt}},
\end{equation}

where $\mathcal{W}(\theta_\mathrm{ker})$ can be a single-Gaussian filter or a starlet filter.  Concerning $\sigma_{n}^\mathrm{filt}$, its definition depends on the employed filter. For a Gaussian kernel, it is given by the standard deviation of the smoothed noise maps, while for the starlet case we need to estimate the noise at each wavelet scale for each image per redshift. To estimate the noise level at each starlet scale, we followed \cite{1998PASP..110..193S} and used the fact that the standard deviation of the noise at the scale $j$ is given by $\sigma_j=\sigma^e_j\sigma_I$, where $\sigma_I$ is the standard deviation of the noise of the image and $\sigma^e_j$ are the coefficients obtained by taking the standard deviation of the starlet transform of a Gaussian distribution with standard deviation one at each scale $j$. For the purposes of this study, we mimicked the shape noise expected for a survey such as \Euclid\footnote{\url{https://www.Euclid-ec.org}}\citep{2011arXiv1110.3193L, 2020A&A...642A.191E} as defined in \autoref{AppendixA}. We performed a tomographic analysis using four source redshifts, $z_s=\left\lbrace 0.5, 1.0, 1.5, 2.0 \right\rbrace$, with corresponding values for the galaxy number density $n_\mathrm{gal}$ per source redshift bin $n_{gal}=\lbrace 11.02, 11.90, 5.45, 1.45\rbrace$.

\begin{figure*}[!htbp]
\includegraphics[width=0.45\textwidth]{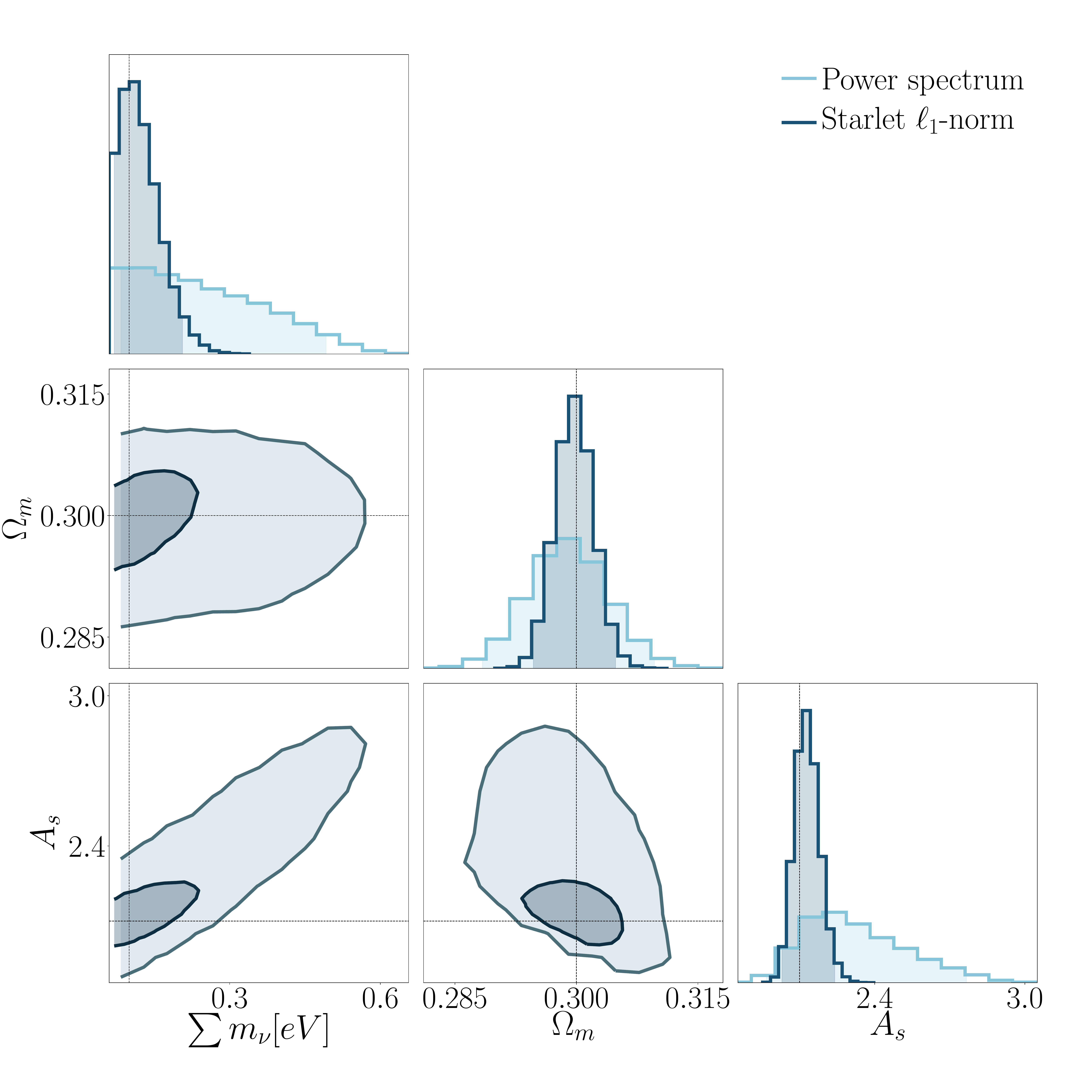}
\includegraphics[width=0.45\textwidth]{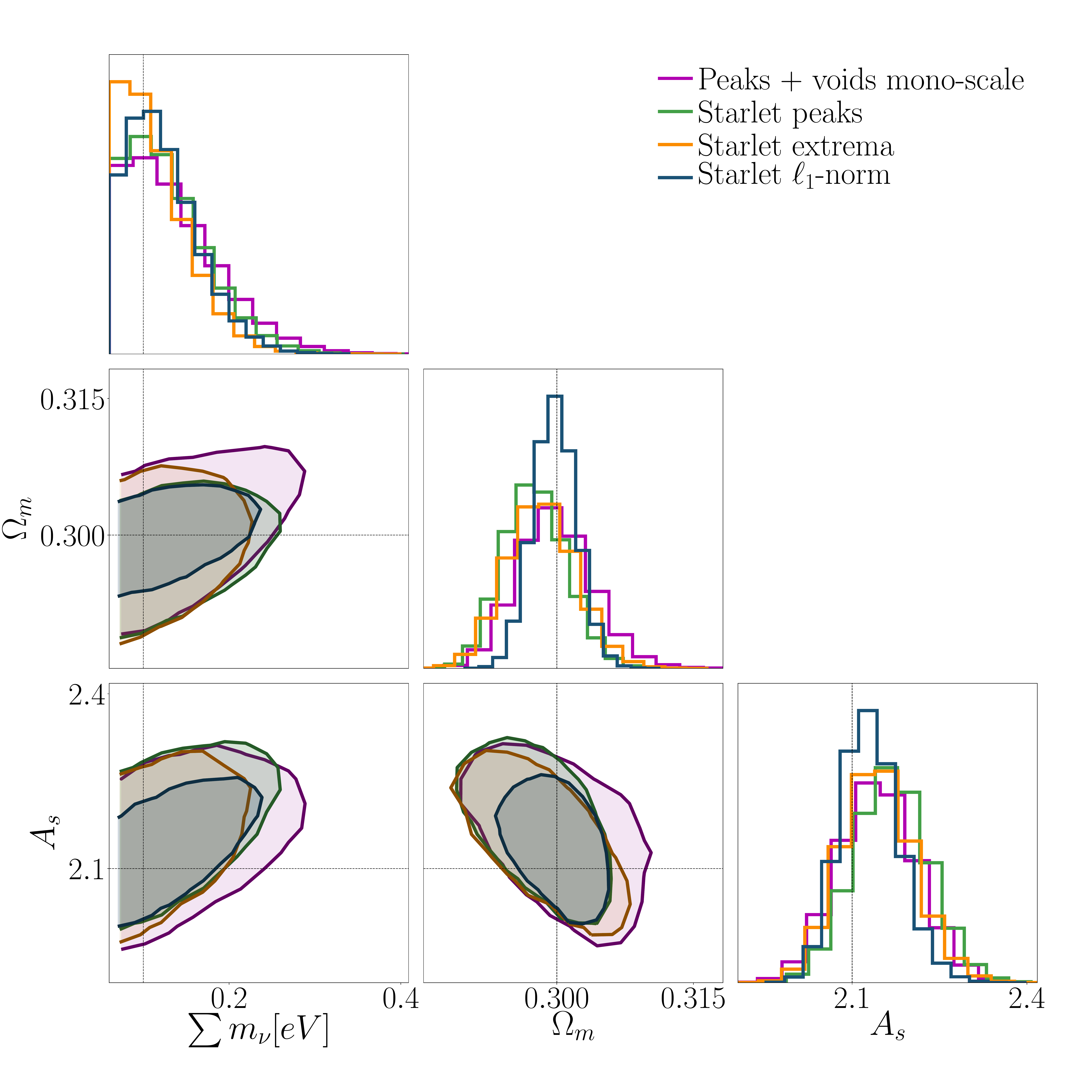}
\caption{95\%  confidence contour tomography with source redshifts $z_s=[0.5,1.0,1.5,2.0]$ and corresponding galaxy number densities $n_\mathrm{gal}=[11.02, 11.90, 5.45, 1.45]$. The dotted black line is the fiducial model: $[\sum m_{\nu}, \Omega_m, 10^9A_s] = [0.1, 0.3, 2.1]$. \textbf{Left panel}: Constraints from the power spectrum (light blue contours) computed on noisy maps smoothed with a Gaussian filter with $\theta_{\rm ker}=  $1 arcmin, compared to constraints from the starlet $\ell_1$-norm (dark blue contours) computed on noisy maps filtered with a four-scale starlet kernel. \textbf{Right panel}: Constraints from the combination of peaks and voids (magenta contours) computed on noisy maps smoothed with a Gaussian filter with $\theta_{\rm ker}= $2 arcmin compared to constraints from starlet peak counts (green contours), starlet extrema (orange contours), and the $\ell_1$-norm (dark blue contours), computed on noisy maps filtered with a four-scale starlet kernel.
\label{fig: power_spectrum_l1_norm_starlet_extrema}}
\end{figure*}

A: We computed the power spectrum as defined in Equation \eqref{eq: power_spectrum} on noisy convergence maps filtered with a Gaussian kernel with smoothing size $\theta_{\ker}$ = 1 arcmin. We considered angular scales with 24 logarithmically spaced bins in the range $\ell=[300,5000]$.

B: Regarding mono-scale peaks and voids, we computed peaks (as pixels with values larger than their eight neighbours) and voids (as pixels with values smaller than their eight neighbours) on noisy convergence maps.\ The maps were filtered with a single-Gaussian kernel with smoothing size $\theta_{\ker}$ = 2 arcmin, with 29 linearly spaced bins for peaks between the minimum and maximum of the map in $S/N$ and 29 linearly spaced bins  for voids between the negative maximum and positive minimum of the maps in $S/N$.
 
   C: Starlet peak counts were computed as pixels with values larger than their eight neighbours. They were computed on noisy convergence maps filtered with a starlet kernel with four scales corresponding to [1.6\arcmin, 3.2\arcmin, 6.4\arcmin, coarse], with 29 linearly spaced bins for each scale between the minimum and  maximum values of each $S/N$ map.

    D: Starlet extrema were obtained by combining the peaks computed on maps with only $S/N>0$ contribution with starlet {voids} computed on maps with only $S/N<0$ contribution.\ We used a starlet as we did with (C), with 58 linearly spaced bins (29 for peaks and 29 for voids).
   
E: The starlet $\ell_1$-norm $S_{\ell_1}$ was computed following Equation \eqref{eq: L1_norm_starlet} on noisy convergence maps filtered with a starlet kernel with four scales. We followed the same process as we did for (C) and (D).\\  
\\
Statistics (D) and (E) are our new proposals. In all statistics where we employed the starlet decomposition, the finest frequency we considered was $\theta_{\rm \ker} = $ 1.6 arcmin, corresponding to the maximum angular scale $\ell_{\rm max} = 2149$.

\setlength{\tabcolsep}{26pt} 
\renewcommand{\arraystretch}{1.1}

\begin{table*}
\caption{\label{tab: Table_FoM}Values of the figure of merit (FoM) as defined in Equation \eqref{eq:  Figure_of_Merit} for different parameter pairs for each observable employed in the likelihood analysis. In the last column, we provide the 3D FoM given as the inverse of the volume in ($M_{\nu}, A_s, \Omega_m$) space.} 
\begin{tabularx}{\textwidth}{lccccc}
\hline
\hline
\textbf{FoM} & $(M_{\nu}$, $\Omega_m$)& ($M_{\nu}$, $A_s$) & ($\Omega_m$, $A_s$) & ($M_{\nu}$, $\Omega_m$, $A_s$) \\
\hline
Power spectrum & 1585 & 77 & 1079 & 2063  \\
Starlet Peaks & 5904 & 331 & 4856 & 10147 \\
$\ell_1$-norm & 11408 & 619 & 9126 & 16688  \\
Starlet extrema & 6967 & 442 & 5956 & 6740  \\
Peaks + voids monoscale & 5321 & 307 & 4286 & 6114\\
\hline
\end{tabularx}
\end{table*}

\subsection{Covariance matrices}

The \url{MassiveNus} suite of simulations (described in more detail in \autoref{AppendixA}) also includes a cosmology with massless neutrinos $ \lbrace {\rm M}_{\nu}, \Omega_{\rm m}, 10^{9}A_{\rm s} \rbrace $=$ \lbrace 0.0, 0.3, 2.1 \rbrace $  obtained from initial conditions different from those of the massive neutrino simulations. We used this additional simulation set to compute the covariance matrix of the observable. The covariance matrix elements are computed as
\begin{equation}\label{covariance_element}
C_{ij}=\sum\limits_{r=1}^{N} \frac{(x_{i}^{r} - \mu_i)(x_{j}^{r} - \mu_j)}{N-1}
,\end{equation}

\noindent where $N$ is the number of observations (in this case, the 10000 realisations), $x_i^{r}$ is the value of the considered observable in the $i^{th}$ bin for a given realisation $r,$ and 
\begin{equation}
\mu_i=\frac{1}{N}\sum_r x_{i}^{r}
\end{equation}

\noindent is the mean of the observable in a given bin over all the realisations. 
Furthermore, we took into account the loss of information due to the finite number of bins and realisations by adopting the estimator introduced by \cite{2007A&A...464..399H} for the inverse of the covariance matrix: 
\begin{equation}\label{Cov_est_corrected}
C^{-1}=\frac{N-n_\mathrm{bins}-2}{N-1}C_{*}^{-1},
\end{equation}

\noindent where $N$ is the number of realisations, $n_\mathrm{bins}$ the number of bins, and $C_{*}$ the covariance matrix computed for the power spectrum and peak counts, whose elements are given by Equation \eqref{covariance_element}. For illustration, we scaled the covariance for a \Euclid \footnote{\url{https://www.euclid-ec.org/}} sky coverage by the factor $f_\mathrm{map}/f_\mathrm{survey}$, where $f_\mathrm{map}=12.25$ deg$^{2}$ is the size of the convergence maps and $f_{Euclid}=15000$ deg$^{2}$.
\label{Section:covariance}

\subsection{Likelihood}

To perform Bayesian inference and get the probability distributions of the cosmological parameters, we used a Gaussian likelihood for a cosmology-independent covariance:
\begin{equation}\label{Likelihood_function}
\log \mathcal{L}(\mathbf{\theta})=\frac{1}{2}(d-\mu(\mathbf{\theta}))^{T}C^{-1}(d-\mu(\mathbf{\theta})),
\end{equation}

\noindent where $d$ is the data array, $C$ is the covariance matrix of the observable, and $\mu$ is the expected theoretical prediction as a function of the cosmological parameters $\theta$. In our case, the data array is the mean over the (simulated) realisations of all the separate observables, or, of their different combinations, for our fiducial model. Cosmological parameters are the ones for which simulations are available, namely $\lbrace {\rm M}_{\nu}, \Omega_{\rm m}, 10^{9}A_{\rm s} \rbrace,$ but the same statistics can of course be applied to other parameters if simulations allow.

In order to predict the theoretical values of all observables given a new set of values for the cosmological parameters $\lbrace {\rm M}_{\nu}, \Omega_{\rm m}, 10^{9}A_{\rm s} \rbrace$, we employed an interpolation with Gaussian process regression \citep{10.5555/1162254} using the \url{scikit-learn} python package. The emulator used in this paper is the same as that employed in \cite{2019PhRvD..99f3527L}.

\subsection{Result estimators}
To quantify the constraining power of the different summary statistics, we employed the following estimators, which are based on \cite{2020A&A...642A.191E}. To have an approximate quantification of the size of the parameter contours, we considered the following figure of merit (FoM):
\begin{equation}
    \text{FoM}=\left ({\det{(\Tilde{F})}}\right)^{1/n}\label{eq:  Figure_of_Merit}
,\end{equation}

\noindent where $\Tilde{F}$ is the marginalised Fisher sub-matrix that we estimated as the inverse of the covariance matrix among the set of cosmological parameters under investigation and  $n$ is equal to the parameter space dimensionality. We show the values of the FoM for our observables in Table \ref{tab: Table_FoM}. To estimate the marginalised 1$\sigma$ error on a single parameter $\theta_{\alpha}$ (which means having included all the degeneracies with respect to other parameters), we used the quantity:\begin{equation}
    \sigma_{\rm \alpha \alpha}=\sqrt{C_{\rm \alpha \alpha}}\label{eq: marginalised_error}
,\end{equation}

\noindent where $C_{\rm \alpha \alpha}$ are the diagonal elements of the parameter covariance matrix.  We show the values of the $\sigma_{\alpha \alpha}$ for our observables in Table \ref{tab: marginalised_errors}.

\subsection{MCMC simulations and posterior distributions}
We explored and constrained the parameter space with the \url{emcee} package, which is a  python implementation of the affine invariant ensemble sampler for Markov chain Monte Carlo (MCMC) introduced by \cite{2013PASP..125..306F}. We assumed a flat prior, a Gaussian likelihood function as defined in Sect. \ref{Likelihood_function}, and a model-independent covariance matrix as discussed in Sect. \ref{Section:covariance}. The walkers were initialised in a tiny Gaussian ball of radius $10^{-3}$ around the fiducial cosmology $[M_{\nu},\Omega_m,10^{9}A_s]=[0.1,0.3,2.1],$ and we estimated the posterior using 120 walkers. Our chains are stable against the length of the chain, and we verified their convergence by employing Gelman Rubin diagnostics \citep{1992StaSc...7..457G}. To plot the contours, we used the \url{ChainConsumer} python package \citep{2016JOSS....1...45H}.

\setlength{\tabcolsep}{10pt} 
\renewcommand{\arraystretch}{1.1}

\begin{table}
\centering
\caption{\label{tab: marginalised_errors}Values of 1-$\sigma$ marginalised errors as defined in Equation \eqref{eq: marginalised_error} for each cosmological parameter for the different observables.}
\begin{tabular}{lccccc}
\hline
\hline
$\mathbf{\sigma_{\alpha\alpha}}$ & $\mathbf{M_{\nu}}$ & $\mathbf{\Omega_m}$ & $\mathbf{A_s}$ \\
\hline
Power spectrum & 0.147 & 0.005 & 0.204\\
Starlet peaks & 0.057 & 0.003 & 0.064\\
$\ell_1$-norm & 0.041 & 0.002 & 0.051\\
Starlet extrema & 0.040 & 0.004 & 0.062\\
Peaks + voids monoscale & 0.054 & 0.004 & 0.067\\

\hline
\end{tabular}
\end{table}

\section{Results}\label{sec: Results}
We now illustrate forecast results on the sum of neutrino masses ${\rm M}_{\nu}$,  on the matter density parameter $\Omega_{\rm m}$, and  on  the power  spectrum  amplitude $A_{\rm s}$ for  a  survey with \Euclid-like noise in a tomographic setting with four source redshifts, $z_s$= [0.5,1.0,1.5,2.0]. We compare results for the  different observables defined in Sect. \ref{subsec: summary_statistics} and investigate the impact of the choice of the filter on the covariance matrix. 

\begin{figure}[h]
    \includegraphics[width = 9 cm]{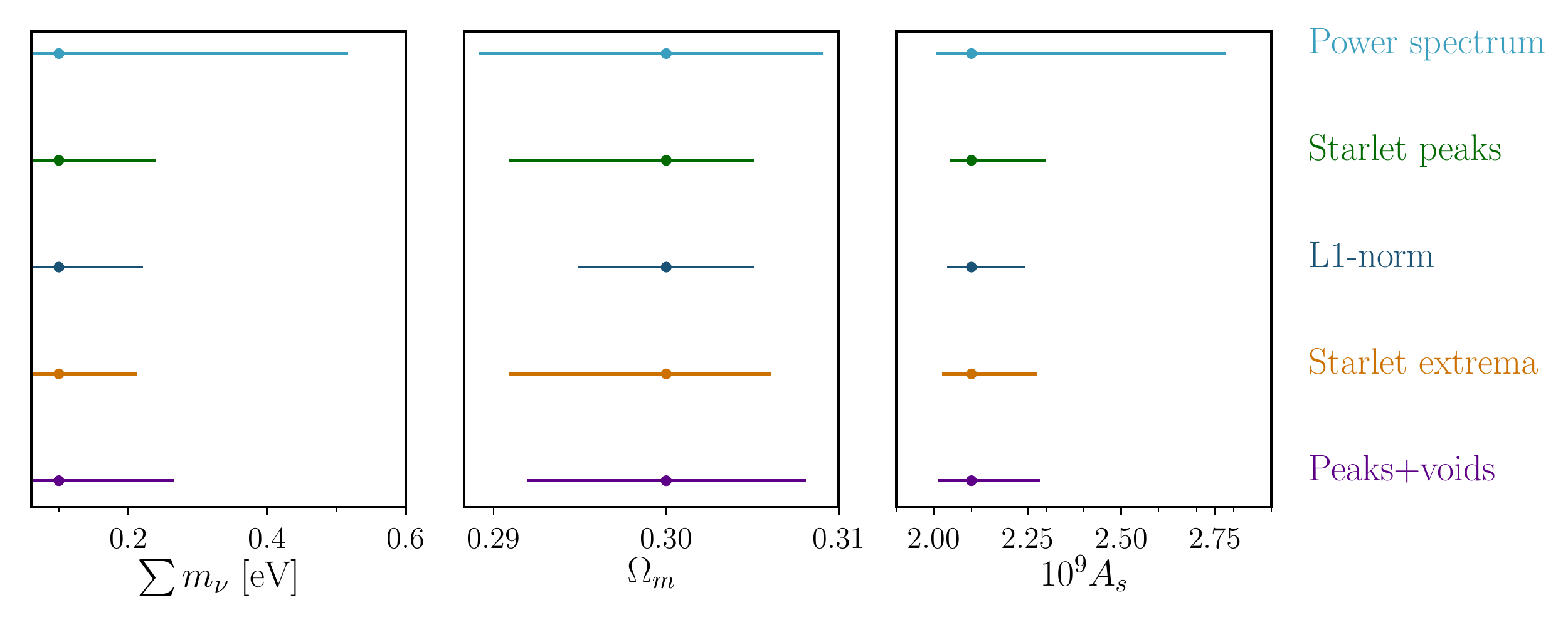}
    \caption{Marginalised errors for the observables described in Sect. \ref{subsec: summary_statistics} on each parameter, showing the 2.5 and 97.5 percentiles with respect to the fiducial model. They refer to a tomographic setting with $z=[0.5, 1.0, 1.5, 2.0]$, with the fiducial model set to $ [M_{\nu}, \Omega_m, 10^{9}A_s]=[ 0.1, 0.3, 2.1 ]$. The last observable refers to mono-scale peaks and voids, as described in the text.}\label{fig:Marginalised_plot}
\end{figure}
In Fig. \ref{fig: power_spectrum_l1_norm_starlet_extrema}, we show the comparison between the constraints obtained using different summary statistics, as described in Sect. \ref{subsec: summary_statistics}. 
As expected, we see that all higher-order statistics are more constraining than the power spectrum. The new result of this letter is represented by the starlet $\ell_1$-norm: The inclusion of all pixels enables us to retrieve tighter constraints than the combination of local minima and maxima (voids+peaks).
Specifically, for all parameter space planes, the $\ell_1$-norm FoM values, illustrated in Table \ref{tab: Table_FoM}, are about twice as large compared to those for the combined local minima and maxima (and  more than seven times larger than the power spectrum FoM values). 

In this work, we have also introduced starlet extrema as a new summary statistic to constrain the parameters. Similarly to the $\ell_1$-norm, starlet extrema are computed between the minimum and maximum $S/N$ value of the map, but they are defined as the combination of local maxima computed on $S/N>0$ and local minima computed on $S/N<0$ (namely, they do not encode the information present in all pixels). We see that starlet extrema FoM values are larger than starlet peaks and peaks+voids mono-scales, suggesting that starlet extrema can be a good multi-scale higher-order statistic candidate. However, the $\ell_1$-norm remains the statistic that performs the best in terms of constraining power with respect to all the summary statistics we have considered. In Fig. \ref{fig:Marginalised_plot}, we show the marginalised constraints on each cosmological parameter corresponding to the different observables. To compare the improvement obtained by employing the different statistics, we computed the 1$\sigma$ marginalised error for each parameter, as summarised in Table \ref{tab: marginalised_errors}. We find that the starlet $\ell_1$-norm outperforms the power spectrum by $72\%$ on M$_{\nu}$, $60\%$ on $\Omega_{\rm m}$, and $75\%$ on $A_s$, and the state-of-the-art peaks+voids for a single smoothing scale by $24\%$ on M$_{\nu}$, $50\%$ on $\Omega_{\rm m}$, and $24\%$ on $A_s$.
Starlet extrema outperform the power spectrum by $72\%$ on M$_{\nu}$, $20\%$ on $\Omega_m$, and $70\%$ on $A_s$.
We also quantify the improvement provided by the $\ell_1$-norm with respect to our previous study \citep{PhysRevD.102.103531}, finding that the starlet $\ell_1$-norm outperforms starlet peaks by $28\%$ on M$_{\nu}$, $33\%$ on $\Omega_m$, and $20\%$ on $A_s$. In \autoref{AppendixB}, we also compare the covariance matrices obtained when using starlet extrema and the $\ell_1$-norm, and we  find that starlet extrema present a more diagonal covariance for the observable.

\section{Conclusions}\label{sec: Conclusions}
In this letter, we have proposed using starlet  $\ell_1$-norm statistics on weak lensing converge maps to constrain cosmological parameters.
The measure of multi-scale peak amplitudes can be seen as a measure of the $\ell_1$-norm of a subset of positive wavelet coefficients. 
Similarly, the measure of void amplitudes  can be seen as a measure of the $\ell_1$-norm of a subset of negative wavelet coefficients. Wavelets therefore provide a great framework for a joint peak and void analysis, in which information from all wavelet coefficients is included. We therefore propose using a very simple $\ell_1$-norm statistic, defined as the sum of the $\ell_1$-norm of all coefficients in a given $S/N$ (pixel) bin for each wavelet scale, as defined in Equation \eqref{eq: L1_norm_starlet}. We investigate the impact of employing the starlet $\ell_1$-norm as summary statistics computed on weak lensing convergence maps to estimate cosmological parameters, and we find that the $\ell_1$-norm outperforms the two state-of-the-art summary statistics, the power spectrum and the combination of mono-scale peaks and voids, by, respectively, $72\%$ and $24\%$ on M$_{\nu}$, $60\%$ and $50\%$ on $\Omega_m$, and $75\%$ and $24\%$ on $A_s$. We have furthered proposed starlet extrema and compared them to the $\ell_1$-norm: In this case as well, the latter performs better in terms of constraining power, within the current ideal setting, while the former present the advantage of a more diagonal covariance matrix. We are aware that the statistical
power alone is not sufficient to serve as a robust probe for precision cosmology; for their usage, it will be important to test how these statistics react in a non-ideal setting and how their performance is impacted by systematics in the signal. We will dedicate a future study to this aspect.

We conclude that the new statistic proposed here presents several advantages in the context of cosmological parameter inference: It provides a fast calculation of the full void and peak distribution; it does not rely on a specific definition of peaks and voids or some arbitrary threshold; it instead encodes information of the entire pixel distribution without excluding pixels that are not local minima or maxima; and it leads to tighter constraints, at least within an ideal setting. The starlet decomposition therefore provides a very powerful framework for a joint multi-scale peak and void analysis.

\begin{acknowledgements}
We thank Jia Liu and the Columbia Lensing group for making the \url{MassiveNus} \citep{2018JCAP...03..049L} simulations available. Virginia Ajani acknowledges support by the Centre National d’Etudes Spatiales and the project Initiative d’Excellence (IdEx) of Université de Paris (ANR-18-IDEX-0001). We thank the Columbia Lensing group\footnote{\href{http://columbialensing.org}{\url{http://columbialensing.org}}} for making their simulations available. The creation of these simulations is supported through grants NSF AST-1210877, NSF AST-140041, and NASA ATP-80NSSC18K1093. We thank New Mexico State University (USA) and Instituto de Astrofisica de Andalucia CSIC (Spain) for hosting the Skies \& Universes site for cosmological simulation product. 
\end{acknowledgements}

\bibliographystyle{aa} 
\bibliography{biblio} 

\begin{appendix}
\section{Background}\label{AppendixA}

\subsection{Weak lensing}\label{subsec:weak_lensing}
The effect of gravitational lensing at co-moving angular distance $f_K(\chi)$ can be described by the lensing potential,

\begin{equation}\label{eq: Lensing_Potential}
\psi(\vec{\theta},\chi)\equiv \frac{2}{c^2}\int_0^{\chi}\mathrm{d}\chi'\frac{f_{K}(\chi - \chi')}{f_K(\chi)f_K(\chi')}\Phi(f_K(\chi')\vec{\theta},\chi') \,\, ,
\end{equation}

\noindent which defines how much the gravitational potential $\Phi$  arising from a mass distribution changes the direction of a light path. In this expression, $K$ is the spatial curvature constant of the universe, $\chi$ is the co-moving radial coordinate, $\vec{\theta}$ is the angle of observation, and $c$ is the speed of light. 
As we are working under the assumption of a Lambda cold dark matter ($\Lambda$CDM) model, the two Bardeen gravitational potentials are here assumed to be equal and the metric signature is defined as $(+1,-1,-1,-1)$.
In particular, under the Born approximation, the effect of the lensing potential on the shapes of background galaxies in the weak regime can be summarised by its variation with respect to $\vec{\theta}$. Formally, this effect can be described by the elements of the lensing potential Jacobi matrix,

\begin{equation}\label{eq: Lensing_Jacobi_Matrix_elements}
A_{ij}=\delta_{ij}-\partial_i \partial_j \psi,
\end{equation}

which can be parametrised as

\begin{equation}\label{eq: Lensing_Jacobi_Matrix}
A=\left (
\begin{array}{cc}
1-\kappa -\gamma_1 &  -\gamma_2   \\
 -\gamma_2  & 1-\kappa +\gamma_1 \\
\end{array}
\right ),
\end{equation}

where $(\gamma_1, \gamma_2)$ are the components of a spin-2 field $\gamma$ called shear and $\kappa$ is a scalar quantity called convergence. They describe, respectively, the anisotropic stretching and the isotropic magnification of the galaxy shape when light passes through large-scale structure. Equations \eqref{eq: Lensing_Jacobi_Matrix_elements} and \eqref{eq: Lensing_Jacobi_Matrix} define the shear and the convergence fields as second-order derivatives of the lensing potential:
\begin{equation}\label{eq: shear}
\gamma_1\equiv\frac{1}{2}(\partial_1 \partial_1 - \partial_2 \partial_2)\psi\quad \gamma_2\equiv \partial_1\partial_2\psi
,\end{equation}
\begin{equation}\label{eq: convergence}
 \kappa\equiv\frac{1}{2}(\partial_1 \partial_1 + \partial_2 \partial_2)\psi=\frac{1}{2}\nabla^2\psi.
\end{equation}
The weak lensing field is a powerful tool for cosmological inference. The shear is more closely related to actual observables (galaxy shapes), while the convergence, as a scalar field, can be more directly understood in terms of the matter density distribution along the line of sight. This can be seen by inserting the lensing potential defined in Equation \eqref{eq: Lensing_Potential} inside Equation \eqref{eq: convergence} and using the fact that the gravitational potential $\Phi$ is related to the matter density contrast $\delta=\Delta\rho/ \bar{\rho}$ through the Poisson equation $\nabla^2 \Phi=4\pi Ga^2 \bar{\rho}\delta$. Expressing the mean matter density in terms of the critical density $\rho_{c,0}=3H_0^2/(8\pi G)$, the convergence field can be rewritten as
\begin{equation}\label{eq: Convergence}
\kappa(\vec{\theta})=\frac{3H_0^2\Omega_m}{2c^2}\int_0^{\chi_\mathrm{lim}}\frac{\mathrm{d}\chi}{a(\chi)}g(\chi)f_{K}(\chi)\delta(f_K(\chi)\vec{\theta},\chi),
\end{equation}
 
\noindent where $H_0$ is the Hubble parameter at its present value and
\begin{equation}
g(\chi) \equiv \int_{\chi}^{\chi_\mathrm{lim}} \mathrm{d}\chi' n(\chi') \frac{f_K(\chi'-\chi)}{f_K(\chi')}
\end{equation}

\noindent is the lensing efficiency.
\noindent Equation \eqref{eq: Convergence} relates the convergence $\kappa$ to the 3D matter over-density field $\delta(f_K(\chi)\vec{\theta},\chi)$, and it describes how the lensing effect on the matter density distribution is quantified by the lensing strength at a distance $\chi$, which directly depends on the normalised source galaxy distribution $n(z)\mathrm{d}z=n(\chi)\mathrm{d}\chi$ and on the geometry of the universe through $f_K(\chi)$ along the line of sight. A complete derivation can be found in \cite{2015RPPh...78h6901K} and \cite{1992grle.book.....S}. 

\subsubsection{Convergence power spectrum}
\noindent To provide a statistical estimate of the distribution of the convergence field, the first non-zero order is given by its second moment, which is commonly described by the `two-point correlation function' ($2$PCF) in real space $\langle \kappa(\theta)\kappa(\theta') \rangle$, or by its counterpart in Fourier space, the `convergence power spectrum':

\begin{equation}
C_{\kappa}(\ell)=\frac{9\Omega_m^2 H_0^4 }{4c^4} \int_0^{\chi_\mathrm{lim}} \mathrm{d}\chi \frac{g^2(\chi)}{a^2(\chi)} P_{\delta} \left(\frac{\ell}{f_{\kappa}(\chi)},\chi \right)\label{eq: power_spectrum}
,\end{equation}

\noindent where $P_{\delta}$ represents the 3D matter power spectrum, directly related to the matter density distribution $\delta$ in Eq. \eqref{eq: Convergence}, of the weak lensing convergence field. In this study, we computed the power spectrum of the noisy filtered convergence maps: For a given cosmology, we added Gaussian noise to each realisation of $\kappa$. 
\noindent To filter the maps, we employed a Gaussian kernel with smoothing size $\theta_\mathrm{ker}=1$ arcmin and considered angular scales with logarithmically spaced bins in the range $\ell=[300,5000]$. We computed the power spectra using \url{LensTools} for each of the 10000 realisations per cosmology, and then we took the average over the realisations. We parallelised our code using \url{joblib}\footnote{\url{https://joblib.readthedocs.io/}} to accelerate processing due to the large number of realisations per cosmology.

\begin{figure*}[!htbp]
\includegraphics[width=0.45\textwidth]{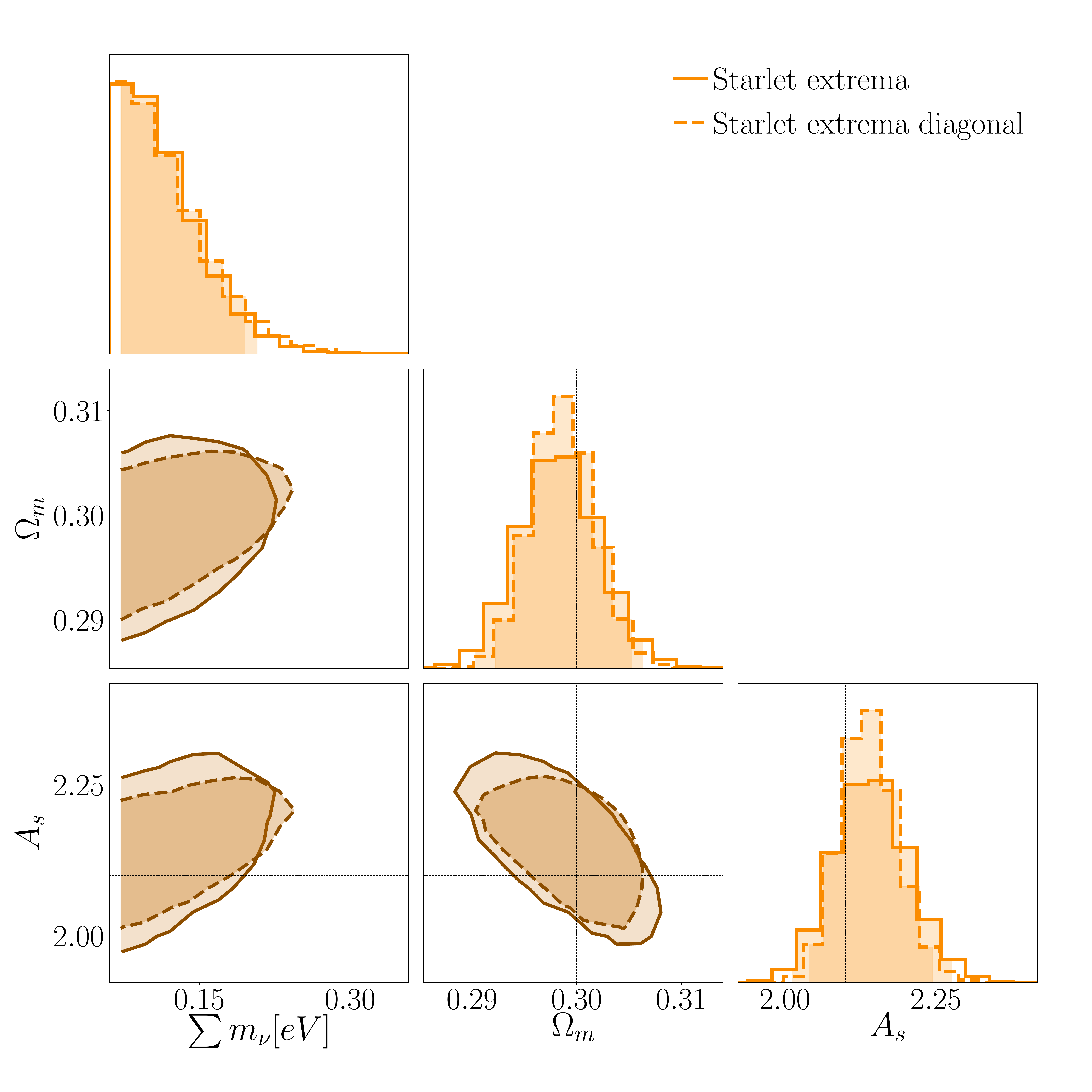}
\includegraphics[width=0.45\textwidth]{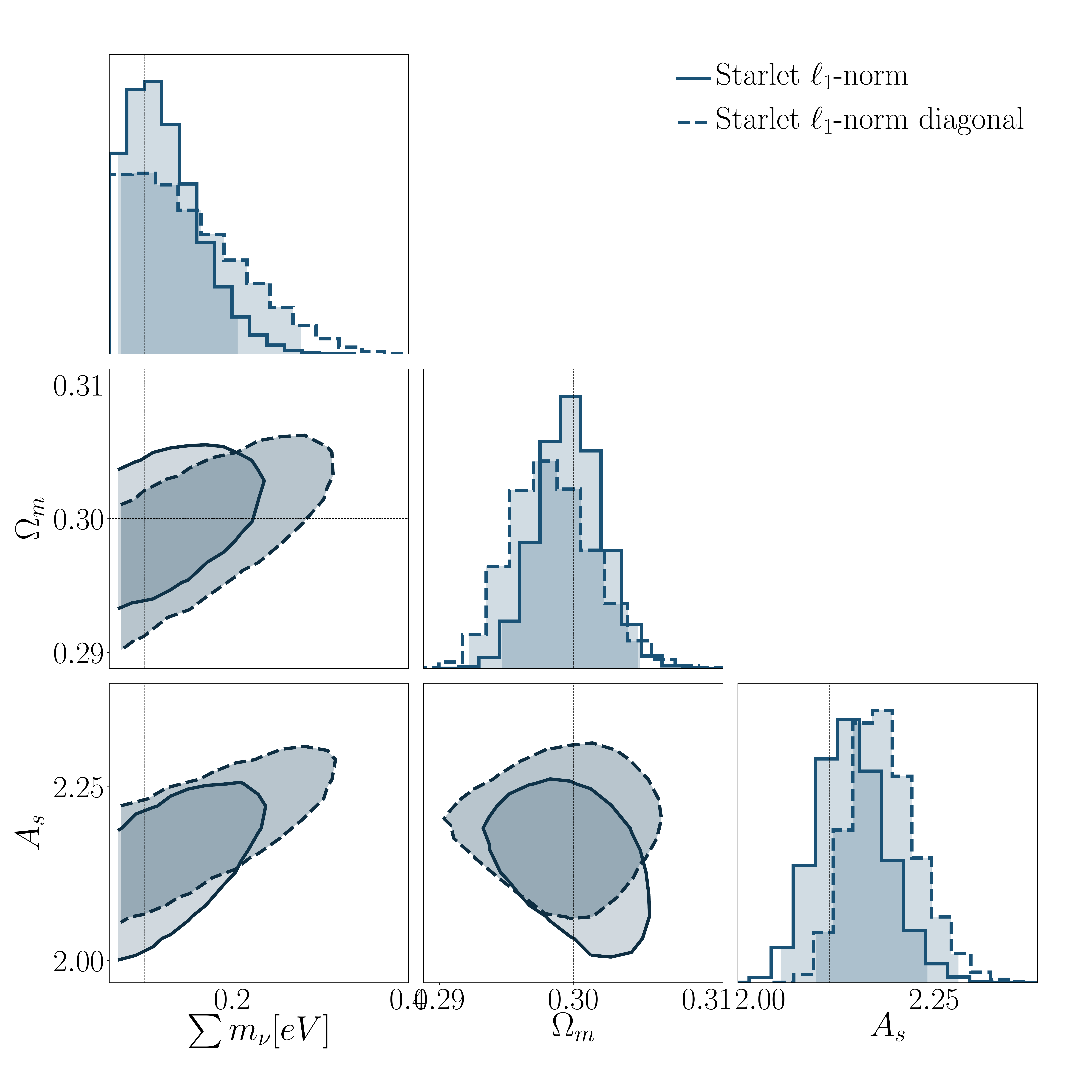}
\caption{95\%  confidence contour tomography with source redshifts $z_s=[0.5,1.0,1.5,2.0]$ and corresponding galaxy number densities: $n_\mathrm{gal}=[11.02, 11.90, 5.45, 1.45]$. The dotted black line is the fiducial model: $[\sum m_{\nu}, \Omega_m, 10^9A_s] = [0.1, 0.3, 2.1]$. \textbf{Left panel}: Constraints from starlet extrema with the full covariance matrix (continuous contours) computed on noisy maps filtered with a four-scale starlet kernel against constraints from starlet extrema with the only diagonal elements of the covariance matrix (dashed contours). \textbf{Right panel}: Constraints from the $\ell_1$-norm with the full covariance matrix (continuous contours) computed on noisy maps filtered with a four-scale starlet kernel against constraints from $\ell_1$-norm with the only diagonal elements of the covariance matrix (dashed contours).}\label{fig: constraints_diag}
\end{figure*}

\subsection{Simulations}\label{subsec:simulations}
In this paper, we used the Cosmological Massive Neutrino Simulations (\url{MassiveNus}), a suite of publicly available N-body simulations released by the Columbia Lensing group\footnote{\url{http://columbialensing.org}}. It contains 101 different cosmological models obtained by varying the sum of neutrino masses $M_{\nu}$, the total matter density parameter $\Omega_m$, and the primordial power spectrum amplitude $A_s$ at the pivot scale $k_0=0.05$ Mpc$^{-1}$  in the ranges 
$M_{\nu}=[0, 0.62]$ eV, $\Omega_m =[$0.18, 0.42$],$ and $A_s\cdot 10^{9}=[$1.29,  2.91$]$. The reduced Hubble constant $h=0.7$, the spectral index $n_s=0.97$, the baryon density parameter $\Omega_b = 0.046,$ and the dark energy equation of state parameter $w=-1$ were kept fixed under the assumption of a flat universe. The fiducial model was set at $ \left\lbrace M_{\nu}, \Omega_m, 10^{9}A_s  \right\rbrace $=$ \left\lbrace 0.1, 0.3, 2.1 \right\rbrace $.
\noindent The complete description of the implementation and the products is illustrated in \cite{2018JCAP...03..049L}. We used simulated convergence maps as mock data for our analysis. When dealing with real data, the actual observable was the shear field, which can be converted into the convergence field following \cite{1993ApJ...404..441K}. We bypassed this step from $\gamma$ to $\kappa$ and worked with the convergence maps that were directly provided as products from \url{MassiveNus}.
The maps were generated using the public ray-tracing package \url{LensTools} \citep{2016A&C....17...73P} for each of the 101 cosmological models at five source redshifts, $z_s=\left\lbrace 0.5, 1.0, 1.5, 2.0, 2.5 \right\rbrace $. Each redshift has 10000 different map realisations obtained by rotating and shifting the spatial planes. Each $\kappa$ map has $512^2$ pixels, corresponding to a $12.25$ deg$^{2}$ total angular size area in the range $\ell \in [100 , 37000]$ with a resolution of 0.4 arcmin.
We mimicked \Euclid-like shape noise at each source redshift assuming Gaussian noise with mean zero and variance:
\begin{equation}\label{eq:noise}
\sigma_{n}^{2}=\frac{\langle \sigma_{\epsilon}^2\rangle}{n_\mathrm{gal}A_\mathrm{pix}} \,\, ,
\end{equation}
\noindent where we set the dispersion of the ellipticity distribution to $\sigma_{\epsilon}=0.3$ and the pixel area is given by $A_\mathrm{pix} \simeq 0.16$ arcmin$^2$.

\section{Diagonal covariances}\label{AppendixB}

As recalled in the introduction, \cite{PhysRevD.102.103531} found that starlet peak counts have the interesting feature that the corresponding covariance matrix is nearly diagonal. Motivated by this, we have tested in this work whether this feature is also maintained for starlet extrema and for the $\ell_1$-norm. Interestingly,
we find that starlet extrema keep this characteristic, while the $\ell_1$-norm show more correlations in the off-diagonal term. This can be seen by looking at Fig. \ref{fig: constraints_diag}: In the left panel, we show the constraints for the starlet extrema when using the full covariance matrix (continuous contours) and compare them with the starlet extrema when using only the diagonal elements of the covariance matrix (dashed contours). Analogously, we show the same comparison for the $\ell_1$-norm in the right panel. When using starlet extrema as summary statistics, there is no loss of information on $M_{\nu}$, and a slight loss of information of 25\% on $\Omega_m$ and 22\% $A_s$  if we employ a covariance matrix with only its diagonal elements in the likelihood analysis. Concerning the $\ell_1$-norm, it is sufficient to look at the contours to notice how the contours with only diagonal terms are considerably different with respect to the contours obtained with the full covariance matrix: They appear shifted and present a different degree of degeneracy for $(\Omega_m, A_s)$. Hence, considering this result along with the result from Sect. \ref{sec: Results}, we conclude that the $\ell_1$-norm outperforms starlet extrema in terms of constraining power when considering the full covariance in the analysis, but presents a less diagonal matrix than starlet extrema. We conclude that, depending on the context, the $\ell_1$-norm could be a convenient choice when the priority is the constraining power, while starlet extrema might be more useful when one is interested in speeding up the analysis or when the covariance matrix can be difficult to invert. 
\end{appendix}

\end{document}